\documentclass[aps,pra,twocolumn,amsfonts,amssymb,amsmath,showpacs,
floatfix,nofootinbib,citesort]{revtex4-2}
\usepackage{mathrsfs}
\usepackage{amsfonts}
\usepackage{amstext}
\usepackage{amsmath}
\usepackage{amssymb}
\usepackage{bm}
\usepackage{CJK}
\usepackage{bbm}
\usepackage[dvips]{graphicx}
\def\qed{\leavevmode\unskip\penalty9999 \hbox{}\nobreak\hfill
	\quad\hbox{\leavevmode  \hbox to.77778em{%
			\hfil\vrule   \vbox to.675em%
			{\hrule width.6em\vfil\hrule}\vrule\hfil}}
	\par\vskip3pt}

\usepackage{amssymb}
\usepackage{graphicx}
\usepackage{graphics}
\usepackage{amsmath}
\usepackage{amsthm}
\usepackage{color}
\usepackage{dsfont}
\usepackage{textcomp}
\definecolor{darkred}  {rgb}{0.5,0,0}
\definecolor{darkblue} {rgb}{0,0,0.5}
\definecolor{darkgreen}{rgb}{0,0.5,0}
\usepackage{hyperref}
\hypersetup{
	pdftitle = {QRT Proposal},
	pdfauthor = {},
	colorlinks = true,
	urlcolor  = blue,         
	linkcolor = red,     
	citecolor = blue,    
	filecolor = darkred       
}
\usepackage{mathtools}
\def\ra{\rangle}
\def\la{\langle}
\def\bb{\mathbb}
\def\ot{\otimes}
\newtheorem{theorem}{Theorem}

\newtheorem{pro}{Proposition}

\newcommand{\bea}{\begin{eqnarray}}
	\newcommand{\eea}{\end{eqnarray}}
\newcommand{\be}{\begin{equation}}
	\newcommand{\ee}{\end{equation}}
\newcommand{\ba}{\begin{equation}\begin{aligned}}
		\newcommand{\ea}{\end{aligned}\end{equation}}

\newcommand{\beax}{\begin{eqnarray*}}
	\newcommand{\eeax}{\end{eqnarray*}}
\newcommand{\bex}{\begin{equation*}}
	\newcommand{\eex}{\end{equation*}}

\newtheorem{definition}{Definition}
\theoremstyle{remark}

\def\be{\begin{equation}}
	\def\ee{\end{equation}}

\newcommand{\mC}{\mathcal{C}}

\newcommand{\mH}{\mathcal{H}}

\newcommand{\mP}{\mathcal{P}}

\newcommand{\mS}{\mathcal{S}}

\newcommand{\lr}{\rangle\langle}

\newcommand{\tr}{{\rm Tr}}

\begin{document}
	

\preprint{APS/123-QED}
\begin{CJK*}{GB}{gbsn}
\title{Partitewise entanglement\\}
		

\author{Yu Guo}		
\email{guoyu3@aliyun.com}
\author{Ning Yang}

\affiliation{School of Mathematical Sciences, Inner Mongolia University, Hohhot, Inner Mongolia 010021, People's Republic of China}

		
\begin{abstract}
It is known that $\rho^{AB}$ as a bipartite reduced state of the three-qubit Greenberger-Horne-Zeilinger~(\text{GHZ}) state is separable, but part $A$ and part $B$ indeed ``share tripartite entanglement'' in the \text{GHZ} state. Namely, whether a state can ``share'' more entanglement is dependent on the global system it lives in. Here we explore such kind of entanglement in any $n$-partite system with arbitrary dimensions, $n\geqslant3$, and call it partitewise entanglement, which includes pairwise entanglement proposed by Dong {\it et al}. [Phys. Rev. A \textbf{110}, 032420(2024)] as a special case. We propose three classes of the partitewise entanglement measures which are based on the genuine entanglement measure, the minimal bipartition, and the minimal distance from the partitewise separable states, respectively. The former two methods are far-ranging since all of them are defined by the reduced function.[Note: If a positive function $h$ satisfies $h\left( \rho^A\right) = E\left( |\psi\lr\psi|^{AB}\right)$ for some bipartite entanglement measure $E$, $h$ is called the reduced function (of $E$).] Consequently, we establish the framework of the resource theory of the partitewise entanglement. In addition, we investigate the partitewise entanglement extensibility and give a measure of such extensibility, from which we find that the maximal partitewise entanglement extension is its purification. Lastly, the relation between this extensibility and the partitewise entanglement is discussed.

\end{abstract}

\maketitle
\end{CJK*}


\section{Introduction}


The characterization of the multipartite entanglement still remains challenging although it has been investigated extensively in the past three decades~\cite{Shi2025prl,Cao2024prl,Gour2013prl,Huber2010prl,Horodecki2009,Guo2024pra,Hong2012pra,Guhne2005njp,Guo2025jpa,Li2024pra,Hong2015pra,Hong2016pra,Huber2013pra,Huber2015pra,Gao2014prl,Hong2021pla,Hong2023epjp,Li2025aqt}. From a theoretical point of view, there are two classes of multipartite entanglement so far: the $k$-entanglement~\cite{Guo2024pra,Hong2012pra,Li2024pra,Hong2015pra,Hong2016pra,Huber2013pra,Huber2015pra,Gao2014prl} and the $k$-partite entanglement~\cite{Guhne2005njp,Guo2025jpa,Hong2021pla,Hong2023epjp,Li2025aqt}. The former one displays how many split subsystems are entangled under partitions of the systems, while the later one focuses on at most how many particles in the global system are entangled but separable from other particles. They are complementary to each other in some sense. Very recently, another kind of entanglement in the three-qubit state was explored, which is termed the pairwise entanglement~\cite{Dong2024pra}.
It is pointed out that although $\rho^{AB}$ as a bipartite reduced state of the three-qubit Greenberger-Horne-Zeilinger (\text{GHZ}) state $|$\text{GHZ}$\ra=\frac{1}{\sqrt{2}}(|000\ra+|111\ra)$ is separable, ``part $A$ and part $B$ are indeed entangled with each other without tracing out part $C$''. Obviously, such an entanglement is not included in the previous two classes.

If an $n$-partite pure state $|\psi\ra^{A_1A_2\cdots A_n}$ cannot be decomposed as $|\psi\ra^X|\psi\ra^Y$ with $A_1$ is included in $X$ and $A_2$ is included in $Y$, then $A_1$ and $A_2$ contain pairwise entanglement~\cite{Dong2024pra}.
The pairwise entanglement between $A$ and $B$ in the three-qubit pure state is quantified by $\sqrt{C^2(AB)+\tau_{ABC}}$~\cite{Dong2024pra}, which is denoted by $\mC_{A'B'}$, and called pairwise concurrence, where $C$ is concurrence~\cite{Rungta2001pra} and $\tau_{ABC}$ is the three-tangle~\cite{Coffman2000pra}. It is clear that $\mC_{A'B'}>0$ whenever $|\psi\ra^{ABC}$ is the \text{GHZ} state.
For the three-qubit pure state, the three-tangle $\tau_{ABC}$ is symmetric, and it is a genuine entanglement measure (but not faithful)~\cite{Coffman2000pra}. But for the higher dimensional case, $\tau_{ABC}$ is not symmetric. If we still use $\sqrt{C^2(AB)+\tau_{ABC}}$ to quantify the pairwise entanglement between $A$ and $B$, it might lead to $\mC_{A'B'}\neq\mC_{B'A'}$ for some states. We thus need to find the nature of such a measure for higher dimensional systems. Going further, along this line, a natural question arises: Is there $k$-partitewise entanglement for $k\geqslant 3$? If so, how can we quantify these entanglement? The main purpose of this paper is to address these issues.

Taking our motivation from the pairwise entanglement in Ref.~\cite{Dong2024pra}, in this paper,
we propose the concept of $k$-partitewise entanglement that is reduced to pairwise entanglement whenever $k=2$. One special case is the genuine partitewise entanglement and another special case is the strong partitewise entanglement. We then present three kinds of $k$-partitewise entanglement measures which also include the genuine $k$-partitewise entanglement measures and the strong $k$-partitewise entanglement measures.
Lastly, we	explore the (strong) extensibility of the $k$-partitewise entanglement. We find out that any mixed state that contains no pure reduced state can be extended into a genuine entangled state, and moreover we give a measure of such an extensibility and discuss the relation between this extensibility and the partitewise entanglement associated with it.

Throughout this paper, we denote by $A_1A_2\cdots A_n$ an $n$-partite finite-dimensional quantum system with the state space $\mH^{A_1A_2\cdots A_n}:=\mathcal{H}^{A_1}\otimes
\mathcal{H}^{A_2}\otimes\cdots\otimes\mathcal{H}^{A_n}$ and by $\mS^{X}$ the set of all density
operators (or called states) acting on $\mH^{X}$. The superscript or subscript $X$ always denotes the corresponding system. For example, the state in $\mS^{X}$ is denoted by $\rho^X$ (or $\rho_X$ sometimes), and is also denoted by $\rho$ for simplicity whenever the associated system $X$ is clear from the context.
$X_1|X_2|\cdots |X_m$ denotes an $m$-partition of $A_1A_2\cdots A_n$ (for instance, partition $AB|C|DE$ is a three-partition of the five-particle system $ABCDE$ with $X_1=AB$, $X_2=C$ and $X_3=DE$) and the vertical bar indicates the split across which the system is regarded as. The case of $m=n$ is the trivial case without any partition. So $m<n$ in general unless otherwise specified.
We denote the rank of a given operator by $r(\cdot)$.


\section{Partitewise entanglement}
	

Observe that any $k$-partite reduced state $\rho^{A_{1}A_{2}\cdots A_{k}}$ of the generalized $n$-qubit \text{GHZ} state
\beax
|\text{\text{GHZ}}_n\ra&=&\frac{1}{\sqrt 2}(|00\cdots 0\ra+|11\cdots 1\ra)
\eeax
is separable, $2\leqslant k <n$, but the subsystem $A_{1}A_{2}\cdots A_{k}$ share the $n$-party genuine entanglement in the global state $|$\text{GHZ}$_n\ra$. Namely, although $\rho^{A_{1}A_{2}\cdots A_{k}}$ is separable, it is contained in an extension system that is not separable. In addition, any $k$-partite reduced state $\varrho^{A_{1}A_{2}\cdots A_{k}}$ of the generalized $n$-qubit $W$ state
\beax
|W_n\ra=\frac{1}{\sqrt n}(|100\cdots 0\ra+|010\cdots0\ra+\cdots|000\cdots1\ra
\eeax
is entangled, $2\leqslant k <n$, and the subsystem ${A_{1}A_{2}\cdots A_{k}}$ also shared the $n$-party entanglement in $|W_n\ra$. That is, the subsystem ${A_{1}A_{2}\cdots A_{k}}$ shared two parts of entanglement: One is the entanglement in $\varrho^{A_{1}A_{2}\cdots A_{k}}$, and the other is the $n$-party genuine entanglement in $|W_n\ra$. We call all the entanglement ``shared'' by $A_{1}A_{2}\cdots A_{k}$ in such sense the $k$-partitewise entanglement of $A_{1}A_{2}\cdots A_{k}$ in $A_1A_2\cdots A_n$. Conversely, for $|\psi\ra^{AB}|\psi\ra^{CD}|\psi\ra^{EFG}$, the state of the subsystem $ACE$, $\rho^{ACE}$, is a product state. In this case, there is no entanglement shared by $ACE$. In addition, the the entanglement shared by $A_{1}A_{2}\cdots A_{k}$ in $|$\text{GHZ}$_n\ra$ is different from that of $|W_n\ra$. For more clarity, we give the following definitions.

\begin{definition}\label{def1}
Let $|\psi\ra$ be a pure state in $\mH^{A_1A_2\cdots A_n}$, $2\leqslant k <n$. If
\bea\label{k-pwsp}
|\psi\ra=|\psi\ra^{X_1}|\psi\ra^{X_2}\cdots|\psi\ra^{X_k}
\eea
for some $k$ partition $X_1|X_2|\cdots|X_k$ such that $A_{i}$ belongs to $X_i$, $1\leqslant i\leqslant k$,
we then say $|\psi\ra$ is $k$-partitewise separable with respect to $A_{1}A_{2}\cdots A_{k}$ (or $A_{1}A_{2}\cdots A_{k}$ is $k$-partitewise separable in $|\psi\ra$). Or else, it is $k$-partitewise entangled ($k$-PWE) with respect to $A_{1}A_{2}\cdots A_{k}$ (or $A_{1}A_{2}\cdots A_{k}$ is $k$-PWE in $|\psi\ra$). Specifically, if $\rho^{A_{1}A_{2}\cdots A_{k}}=\tr_{A_{k+1}\cdots A_n}|\psi\ra\la\psi|$ is genuinely entangled, $k\geqslant 3$, we say $|\psi\ra$ is genuinely $k$-partitewise entangled (G$k$-PWE) with respect to $A_{1}A_{2}\cdots A_{k}$ (or $A_{1}A_{2}\cdots A_{k}$ is G$k$-PWE in $|\psi\ra$).
\end{definition}

For example, the subsystem $A_{1}A_{2}\cdots A_{k}$ is $k$-PWE but not G$k$-PWE in $|$\text{GHZ}$_n\ra$, while the subsystem $A_{1}A_{2}\cdots A_{k}$ is not only $k$-PWE but also G$k$-PWE in $|W_n\ra$, $3\leqslant k<n$. If
\bea\label{case3}
|\phi\ra=|\phi\ra^{A_{1}A_{2}\cdots A_{k}}|\phi\ra^{A_{k+1}\cdots A_n},
\eea
then $|\phi\ra$ is $k$-PWE (respectively, G$k$-PWE) with respect to $A_{1}A_{2}\cdots A_{k}$ iff $|\phi\ra^{A_{1}A_{2}\cdots A_{k}}$ is $k$ entangled (respectively, genuinely entangled). Here, we say $|\psi\ra\in\mH^{A_1A_2\cdots A_n}$ is $k$ entangled if it is not $k$ separable, where $|\psi\ra$ is called $k$-separable if it admits the form as Eq.~\eqref{k-pwsp}, and also called fully separable if $k=n$.

The case of $k=2$ is reduced to the pairwise entanglement in Ref.~\cite{Dong2024pra}.	
By definition, if $A_{1}A_{2}\cdots A_{k}$ is $k$-partitewise separable in $|\psi\ra$, then $\rho^{A_{1}A_{2}\cdots A_{k}}$ is a product state, i.e., $\rho^{A_{1}A_{2}\cdots A_{k}}=\bigotimes_{i=1}^k\rho^{A_{i}}$. But the converse is not true necessarily. For example, assuming that $\mH^B$ has a subspace isomorphic to $\mH^{B_1}\otimes\mH^{B_2}$, up to local unitary on system $B_1B_2$, we take
\bea
|\psi\ra^{ABC}=|\psi\ra^{AB_1}|\psi\ra^{B_2C}.
\eea
It is straightforward that $\rho^{AC}=\rho^A\ot\rho^C$ but $AC$ is pairwise entangled in $|\psi\ra^{ABC}$ whenever both $|\psi\ra^{AB_1}$ and $|\psi\ra^{B_2C}$ are entangled. Also, we can define the $k$-partitewise entanglement for mixed states as follows.

\begin{definition}
Let $\rho$ be a mixed state in $\mS^{A_1A_2\cdots A_n}$, $2\leqslant k <n$. If
\bea\label{k-pwsm}
\rho=\sum_ip_i|\psi_i\ra\la\psi_i|
\eea
for some ensemble $\{p_i, |\psi_i\ra\}$ where $|\psi_i\ra$s are $k$-partitewise separable with respect to $A_{1}A_{2}\cdots A_{k}$, we say $\rho$ is $k$-partitewise separable with respect to $A_{1}A_{2}\cdots A_{k}$ (or $A_{1}A_{2}\cdots A_{k}$ is $k$-partitewise separable in $\rho$). Or else, it is $k$-PWE with respect to $A_{1}A_{2}\cdots A_{k}$ (or $A_{1}A_{2}\cdots A_{k}$ is $k$-PWE in $\rho$).
\end{definition}

The G$k$-PWE mixed state follows directly. Namely, if $\rho^{A_{1}A_{2}\cdots A_{k}}=\tr_{A_{k+1}\cdots A_n}\rho$ is genuinely entangled, we say $\rho$ is G$k$-PWE with respect to $A_{1}A_{2}\cdots A_{k}$ (or $A_{1}A_{2}\cdots A_{k}$ is G$k$-PWE in $\rho$). Note here that, in Eq.~\eqref{k-pwsm}, the $k$-partition corresponding to $|\psi_i\ra$ might be different from that of $|\psi_j\ra$ when $i\neq j$. For example, if $\rho^{ABCDEF}$ is three-partitewise separable with respect to $ADF$, it may occur that \beax
|\psi_1\ra=|\psi_1\ra^{AB}|\psi_1\ra^{CDE}|\psi_1\ra^{F},& \\
|\psi_2\ra=|\psi_2\ra^{A}|\psi_2\ra^{BCD}|\psi_2\ra^{EF},& \\
|\psi_3\ra=|\psi_3\ra^{ABC}|\psi_3\ra^{D}|\psi_3\ra^{EF},&\\
\cdots.&
\eeax
We quickly observe that any $k$-partitewise separable state is $k$ separable, but the converse is not true in general. By definition, together with Theorem 2 in Ref.~\cite{Guo2020pra}, we have the following criteria for detection of the $k$-partitewise entanglement.

\begin{pro}\label{pro2-1}
If $|\psi\ra\in\mH^{A_1A_2\cdots A_n}$ is $k$-partitewise separable with respect to $A_{1}A_{2}\cdots A_{k}$, $2\leqslant k <n$, then $\rho^{A_{1}A_{2}\cdots A_{k}}$ is a product state. If $\rho\in\mS^{A_1A_2\cdots A_n}$ is $k$-partitewise separable with respect to $A_{1}A_{2}\cdots A_{k}$, then $\rho^{A_{1}A_{2}\cdots A_{k}}$ is fully separable. In particular, if $\dim\mH^{A_i}\leqslant3$ for any $i$,  then $|\psi\ra\in\mH^{A_1A_2\cdots A_n}$ is $k$-partitewise separable with respect to $A_{1}A_{2}\cdots A_{k}$ iff $\rho^{A_{1}A_{2}\cdots A_{k}}$ is a product state, and $\rho\in\mS^{A_1A_2\cdots A_n}$ is $k$-partitewise separable with respect to $A_{1}A_{2}\cdots A_{k}$ iff $\rho^{A_{1}A_{2}\cdots A_{k}}$ is fully separable.
\end{pro}

If $|\psi\ra^{A_1A_2\cdots A_k}$ in Eq.~\eqref{case3} is neither fully separable nor genuinely entangled, the $k$-partitewise entanglement therein is different from both that of $|$\text{GHZ}$_n\ra$ and $|W_n\ra$, i.e., there is no larger genuine entangled system contains the considered reference subsystem $A_{1}A_{2}\cdots A_{k}$. In order to distinguish these different $k$-partitewise entanglement more distinctly, we give the following definition.

\begin{definition}\label{Sk-PWE}
	Let $|\psi\ra$ be a pure state in $\mH^{A_1A_2\cdots A_n}$, $2\leqslant k <n$. If it is
	$k$-PWE with respect to $A_{1}A_{2}\cdots A_{k}$ and, moreover, there exists a partition $X|Y$ such that
	\bea \label{sk-pwe}
	|\psi\ra=|\psi\ra^X|\psi\ra^Y
	\eea
	with $A_1A_2\cdots A_k\subseteq X$ and $|\psi\ra^X$ is genuinely entangled,
	we say $|\psi\ra$ is strongly $k$-partitewise entangled (S$k$-PWE) with respect to $A_{1}A_{2}\cdots A_{k}$ (or $A_{1}A_{2}\cdots A_{k}$ is S$k$-PWE in $|\psi\ra$).
\end{definition}

That is, both $|$\text{GHZ}$_n\ra$ and $|W_n\ra$ are S$k$-PWE with respect to $A_{1}A_{2}\cdots A_{k}$ but $|\phi\ra$ in Eq.~\eqref{case3} is not when $|\psi\ra^{A_1A_2\cdots A_k}$ is not genuinely entangled, $2\leqslant k<n$. In particular, if $|\psi\ra^X$ in Eq.~\eqref{sk-pwe} coincides with $|\psi\ra^{A_1A_1\cdots A_k}$, then $|\psi\ra$ is both G$k$-PWE and S$k$-PWE with respect to $A_{1}A_{2}\cdots A_{k}$. But if $X=A_1A_2\cdots A_kA_{k+1}\cdots A_{k+s}$, $s\geqslant1$, and $\rho^{A_1A_2\cdots A_k}$ is not genuinely entangled, then $|\psi\ra$ is S$k$-PWE but not G$k$-PWE with respect to $A_{1}A_{2}\cdots A_{k}$ (e.g., $|\psi\ra^{X}$ is the \text{GHZ} state). Let
\beax
|\varphi\ra=|\varphi\ra^{AB}|\varphi\ra^{CDE}|\varphi\ra^{FG}|\varphi\ra^{HIG},
\eeax
where $|\varphi\ra^{AB}$ and $|\varphi\ra^{FG}$ are entangled, and $|\varphi\ra^{CDE}$ is genuinely entangled. Then $|\varphi\ra$ is $4$-PWE with respect to $ACDF$, but it is neither G$4$-PWE nor S$4$-PWE with respect to $ACDF$. The S$k$-PWE mixed state can be defined straightforwardly. Hereafter, we only say a state is $k$-partitewise separable or entangled (or $A_{1}A_{2}\cdots A_{k}$ is $k$-partitewise separable or entangled) if the reference subsystem $A_{1}A_{2}\cdots A_{k}$ and the global state $|\psi\ra$ are clear from the context. The G$k$-PWE state and the S$k$-PWE state follow in the same way.

By definition, the pairwise entangled $|\psi\ra\in\mH^{ABC}$ with respect to $AB$ has three different cases: (i) $\rho^{AB}$ is entangled but $|\psi\ra$ is not strongly pairwise entangled, (ii) $\rho^{AB}$ is entangled and $|\psi\ra$ is strongly pairwise entangled, and (iii) $\rho^{AB}$ is separable but $|\psi\ra$ is strongly pairwise entangled. But the $k$-partitewise entangled states can be divided into four classes ($k\geqslant3$): (i) Neither genuinely $k$-partitewise entangled nor strongly $k$-partitewise entangled; (ii) Genuinely $k$-partitewise entangled but not strongly $k$-partitewise entangled; (iii) Strongly $k$-partitewise entangled but not genuinely $k$-partitewise entangled; and (iv) Not only genuinely $k$-partitewise entangled but also strongly $k$-partitewise entangled.


\section{Partitewise entanglement measure}


In this section, we discuss how can we quantify the $k$-partitewise entanglement. We begin with the definition of the pairwise entanglement measure (PEM) that was proposed in Ref.~\cite{Dong2024pra} for the first time. A function $\check{E}^{A_1A_2}:\mS^{A_1A_2\cdots A_n}\longrightarrow\bb{R}^+$ is called a PEM with respect to $A_1A_2$ if
(PE1) $\check{E}^{A_1A_2}(\rho)=0$ iff $\rho$ is pairwise separable with respect to $A_1A_2$;
(PE2) $\check{E}^{A_1A_2}$ does not increase under any $n$-partite local operations and classical communication (LOCC), i.e., $\check{E}^{A_1A_2}[\varepsilon(\rho)]\leqslant  \check{E}^{A_1A_2}(\rho)$ for any $n$-partite LOCC $\varepsilon$. If $\check{E}^{A_1A_2}:\mS^{A_1A_2\cdots A_n}\longrightarrow\bb{R}^+$ satisfies (PE2) and
(PE1$'$) $\check{E}^{A_1A_2}(\rho)=0$ if $\rho$ is pairwise separable with respect to $A_1A_2$,
it is also called a PEM, but it might be nonfaithful. If $\check{E}^{A_1A_2}$ satisfies (PE1) [sometimes (PE1$'$)], it is convex and does not increase under LOCC on average additionally, we call it a pairwise entanglement monotone (PEMo). Throughout this paper, for a measure of any kind of entanglement, if it is convex and does not increase on average under LOCC, it is called a monotone of this kind of entanglement, and we denote by *Mo the latter one, if the former one by *M (* denotes the associated kind of entanglement), e.g., PEMo and PEM is such a case. In addition, as that of (PE1) and (PE1$'$), the former one corresponding to the measure is faithful while the later one is not necessarily faithful.
Straightforwardly, any *Mo must be a *M. The measure of the $k$-paritewise entanglement can now be defined naturally.

\begin{definition}\label{kPWEM}
A function $\check{E}^{A_{1}A_{2}\cdots A_{k}}:\mS^{A_1A_2\cdots A_n}\longrightarrow\bb{R}^+$ ($k\geqslant 2$) is called a $k$-partitewise entanglement measure ($k$-PWEM) with respect to $A_{1}A_{2}\cdots A_{k}$ if it admits
{\rm($k$PWE1)} $\check{E}^{A_{1}A_{2}\cdots A_{k}}(\rho)=0$ iff $\rho$ is $k$-partitewise separable with respect to $A_{1}A_{2}\cdots A_{k}$ and {\rm ($k$PWE2)} $\check{E}^{A_{1}A_{2}\cdots A_{k}}$ does not increase under any $n$-partite LOCC.
\end{definition}

The case of $k=2$ is just the PEM. If a state $\rho$ is $k$-PWE, it is neither necessarily G$k$-PWE, nor necessarily S$k$-PWE in general. We thus give the measures of genuine $k$-partitewise entanglement and strong $k$-partitewise entanglement respectively in the following.

\begin{definition}\label{GkPWEM}
A function $\check{E}_g^{A_{1}A_{2}\cdots A_{k}}:\mS^{A_1A_2\cdots A_n}\longrightarrow\bb{R}^+$ ($k\geqslant 3$) is called a genuine $k$-partitewise entanglement measure (G$k$-PWEM) with respect to $A_{1}A_{2}\cdots A_{k}$ if it admits
{\rm(G$k$PWE1)} $\check{E}_g^{A_{1}A_{2}\cdots A_{k}}(\rho)>0$ iff $\rho^{A_1A_2\cdots A_k}$ is genuinely entangled and
{\rm (G$k$PWE2)} $\check{E}_g^{A_{1}A_{2}\cdots A_{k}}$ does not increase under any $n$-partite LOCC.
\end{definition}

\begin{definition}\label{SkPWEM}
	A function $\check{E}_s^{A_{1}A_{2}\cdots A_{k}}:\mS^{A_1A_2\cdots A_n}\longrightarrow\bb{R}^+$ is called a strong $k$-partitewise entanglement measure (S$k$-PWEM) with respect to $A_{1}A_{2}\cdots A_{k}$ if it admits 
	{\rm(S$k$PWE1)} $\check{E}_s^{A_{1}A_{2}\cdots A_{k}}(\rho)>0$ iff $\rho$ is S$k$-PWE and
	{\rm (S$k$PWE2)} $\check{E}_s^{A_{1}A_{2}\cdots A_{k}}$ does not increase under any $n$-partite LOCC.
\end{definition}

In the three subsections below, we will present three classes of $k$-PWEMs, which are derived from the genuine entanglement monotone (GEMo), the minimal bipartition, and the minimal distance from the partitewise separable states, respectively.

\subsection{$k$-PWEM from GEM}

We begin with the PEM, and then consider $k$-PWEM with $k\geqslant 3$.
By definition, the pairwise entanglement of $AB$ in a tripartite state $|\psi\ra^{ABC}$ is either contained only in $\rho^{AB}$ (e.g., $|\psi\ra^{ABC}=|\psi\ra^{AB}|\psi\ra^C$), or contained only in $|\psi\ra^{ABC}$ (e.g., the \text{GHZ} state), or contained in both $\rho^{AB}$ and $|\psi\ra^{ABC}$ [e.g., the $W$ state $|W\ra=\frac{1}{\sqrt3}(|100\ra++|010\ra+|001\ra)$].
For any reduced function $h$, let $E$ be the associated entanglement measure, and $E_g^{(3)}$ be a genuine entanglement measure induced from $h$~\cite{Suppl}; we define
\bea\label{pem}
\check{E}^{AB}(|\psi\ra^{ABC})=E(AB)+E_g^{(3)}(ABC).
\eea
Equation~\eqref{pem} can be extended into mixed states as
\bea\label{pem'}
\check{E}^{AB}\left(\rho^{ABC}\right)\equiv\min\sum_{j=1}^{n}p_j\check{E}^{AB}\left(|\psi_j\lr\psi_j|^{ABC}\right),
\eea
which is called the convex-roof extension of $\check{E}^{AB}$ in Eq.~\eqref{pem}, where the minimum is taken over all pure state decompositions of $\rho^{ABC}=\sum_{j=1}^{n}p_j|\psi_j\lr\psi_j|^{ABC}$. Hereafter, when we present measure for pure states, the case of mixed state are all defined by the convex-roof extension with no further statement. Obviously, any convex-roof extended measure is convex. It is straightforward that $\check{E}^{AB}$ above is a faithful PEMo if both $E$ and $E_g^{(3)}$ are faithful.

  $\check{E}^{AB}$ quantifies all the entanglement shared by subsystem $AB$ in the global system, where the first part $E(AB)$ quantifies the shared entanglement between $A$ and $B$ independent of the global system while the second part $E_g^{(3)}(ABC)$ reveals the entanglement shared simultaneously with the third subsystem $C$. In nature, the pairwise concurrence is such a case since $\mC_{A'B'}^2$ is just a special case of $\check{E}^{AB}$. But $\tau_{ABC}$ vanished on the $W$ state, so $\mC_{A'B'}$ cannot measure all the PE faithfully.

We now discuss the PE in $|\psi\ra=|\psi\ra^{ABCD}$ with respect to $AB$. If $|\psi\ra=|\psi\ra^{AB}|\psi\ra^{CD}$, then the PE is only contained in $AB$; if $|\psi\ra=|\psi\ra^{ABC}|\psi\ra^{D}$, the PE may be contained in both $AB$ and $ABC$; if $|\psi'\ra=|\psi\ra^{ABD}|\psi\ra^{C}$, the PE may be contained in both $AB$ and $ABD$ (here, $|\psi'\ra$ denotes $|\psi\ra$ up to permutation of the subsystems); if $|\psi\ra^{ABCD}$ is genuinely entangled, the PE may be contained in both $AB$ and $ABCD$. Thus,
for $|\psi\ra=|\psi\ra^{A_1A_2\cdots A_n}\in\mH^{A_1A_2\cdots A_n}$,
Eq.~\eqref{pem} can be extended as
\begin{widetext}
\bea\label{pwem}
\!\!\check{E}^{A_1A_2}(|\psi\ra)=
	\begin{cases}
		E(A_1A_2)+ E_g^{(l)}(A_1A_2A_{3'}\cdots A_{l'}),& \!\!\!\text{$\exists$ $|\psi\ra^{A_1A_2A_{3'}\cdots A_{l'}}
		(3\le l\le n)$ is a genuinely entangled state} \\
		E(A_1A_2), & \!\!\!\text{otherwise},
	\end{cases}
\eea
\end{widetext}
where $E_g^{(l)}$ is a GEM, $E$ and $E_g^{(l)}$ are induced from the same reduced function $h$, $\{3'$, $4'$, $\dots$, $l'\}\subseteq\{3, 4, \dots, n\}$.

For the general case of $k\geqslant3$, for example, we take $|\psi\ra=|\psi\ra^A|\psi\ra^B|\psi\ra^{CDE}$ where $|\psi\ra^{CDE}$ is genuinely entangled; then the four-partitewise entanglement with respect to $ABCD$ is only contained in $|\psi\ra^{CDE}$. If $|\psi\ra=|\psi\ra^{ABCD}|\psi\ra^E$, its four-partitewise entanglement with respect to $ABCD$ is
contained in $|\psi\ra^{ABCD}$. If $|\psi\ra=|\psi\ra^{AB}|\psi\ra^{CD}|\psi\ra^E$, it is contained in $|\psi\ra^{AB}$ and $|\psi\ra^{CD}$, but it is not genuinely four-partitewise entangled. In view of this, we put forward the following candidates of the $k$-PWEM. We fix some notations at first for convenience. If $|\psi\ra=|\psi\ra^{X_1}|\psi\ra^{X_2}\cdots|\psi\ra^{X_l}$ where $|\psi\ra^{X_i}$ is either a single state or a composite non-biseparable state, $1\leqslant l<k$, $1\leqslant i\leqslant l$, we let
$\varUpsilon^{X_1X_2\cdots X_l}=\{X_s: X_s$~contains at least two subsystems of $A_1A_2\cdots A_k$ and also at least one subsystem of $A_{k+1}\cdots A_n\}$
and denote by $t_s$ the number of subsystems contained in $X_s$, $1\leqslant s\leqslant l$. {Hereafter, if we write $|\psi\ra=|\psi\ra^{X_1}|\psi\ra^{X_2}\cdots|\psi\ra^{X_l}$, it always refers to $|\psi\ra^{X_i}$ as either a single state or a composite non-biseparable state. For any $|\psi\ra\in\mH^{A_1A_2\cdots A_n}$, $k\geqslant 3$, we let
\begin{widetext}
\bea \label{kpwem}
\!\!\check{E}^{A_1A_2\cdots A_k}(|\psi\ra)=
\begin{cases}
	E^{(k)}(A_1A_2\cdots A_k)+\sum\limits_{X_s\in\varUpsilon^{X_1X_2\cdots X_l}}E_g^{(t_s)}(|\psi\ra^{X_s}), &\!\! |\psi\ra=|\psi\ra^{X_1}|\psi\ra^{X_2}\cdots|\psi\ra^{X_l}|\psi\ra^Y, ~l<k\\
	E^{(k)}(A_1A_2\cdots A_k), &\!\!\text{otherwise},
\end{cases}
\eea
\end{widetext}
\bea \label{gkpwem}
\check{E}_g^{A_1A_2\cdots A_k}(|\psi\ra)=
	E_g^{(k)}(\rho^{A_1A_2\cdots A_k}),
\eea
and
\bea \label{skpwem}
\check{E}_s^{A_1A_2\cdots A_k}(|\psi\ra)=
\begin{cases}
	E_g^{(t_X)}(|\psi\ra^{X}), & |\psi\ra=|\psi\ra^{X}|\psi\ra^Y ~~~~~\\
	0, &\text{otherwise},
\end{cases}
\eea
where $X_i$ contains at least one subsystem of $A_1A_2\cdots A_k$, $A_1A_2\cdots A_k\subseteq X$, $|\psi\ra^X$ is genuinely entangled, $t_X$ denotes the number of subsystems contained in $X$, $E^{(k)}$ is some multipartite entanglement measure as in Ref.~\cite{Guo2020pra,Guo2024rp}, $E^{(k)}$ and $E_g^{(l)}$ are induced from the same reduced function $h$. For example, we can take~\cite{Guo2020pra,Guo2024rp}
\beax E^{(k)}(|\psi\ra^{A_1A_2\cdots A_k})=\frac12\sum_{i=1}^kS(\rho^{A_i})
\eeax
and~\cite{Guo2024rp}
\beax
E_g^{(l)}(|\psi\ra^{A_1A_2\cdots A_l})=\min\limits_{Z|\overline{Z}} S(\rho^{Z}),
\eeax
where the minimization runs over all possible bipartition $Z|\overline{Z}$ of $A_1A_2\cdots A_l$, $S(\rho)=-\tr(\rho\log_2\rho)$ is the von Neumann entropy, or~\cite{Guo2022entropy}
\beax
E_g^{(l)}(|\psi\ra^{A_1A_2\cdots A_l})=\frac12\delta(|\psi\ra^{A_1A_2\cdots A_l})\sum_{i=1}^{l}S(\rho^{A_i}),
\eeax
where $\delta(|\psi\ra^{A_1A_2\cdots A_l})=1$ when $|\psi\ra^{A_1A_2\cdots A_l}$ is genuinely entangled, and  $\delta(|\psi\ra^{A_1A_2\cdots A_l})=0$ when $|\psi\ra^{A_1A_2\cdots A_l}$ is not genuinely entangled. $E^{(k)}$ can also be any $k$-entanglement measure as in Ref.~\cite{Hong2012pra,Li2024pra,Guo2024pra}. It is easy to check that $\check{E}^{A_1A_2\cdots A_k}$ is a $k$-PWEM, $\check{E}_g^{A_1A_2\cdots A_k}$ is a G$k$-PWEM, and $\check{E}_s^{A_1A_2\cdots A_k}$ is a S$k$-PWEM (they are all $k$-PWEMos if both $E^{(k)}$ and $E_g^{(l)}$ therein are entanglement monotones). For the case of $k=2$, Eq.~\eqref{kpwem} turns back to Eq.~\eqref{pwem}.

\subsection{$k$-PWEMo from 2-partitions}

Considering $|\psi\ra^{ABC}$, if it is pairwise separable with respect to $AB$, then either $E(A|BC)=0$ or $E(B|AC)=0$. Conversely, if $E(A|BC)>0$ and $E(B|AC)>0$, it is pairwise entangled.
Let $h$ be a reduced function, we define
\beax
\check{E}_{\min}^{AB}(|\psi\ra^{ABC})&=&\min\{h(\rho^A), h(\rho^B)\},\\
\check{\bar{E}}^{AB}(|\psi\ra^{ABC})&=&\begin{cases}
	\frac12(h(\rho^A)+h(\rho^B)), &h(\rho^A)h(\rho^B)>0\\
	0,&h(\rho^A)h(\rho^B)=0,
\end{cases}\\
\check{E}_{G}^{AB}(|\psi\ra^{ABC})&=&{[h(\rho^A)h(\rho^B)]}^{\frac12}.
\eeax
For instance, if we take $h_\tau$ as the reduced function of tangle $\tau$~\cite{Rungta2003pra} [$\tau(|\psi\ra)=C^2(|\psi\ra)$ for pure state], i.e., $h_\tau(\rho)=2(1-\tr\rho^2)$, then
\beax
\check{\tau}_{\min}^{AB}(|\psi\ra^{ABC})&=&\min\{h_\tau(\rho^A), h_\tau(\rho^B)\},\\
\check{\bar{\tau}}^{AB}(|\psi\ra^{ABC})&=&\begin{cases}
	\frac12[h_\tau(\rho^A)+h_\tau(\rho^B)], &\!\!\!\tr\rho_{A,B}^2<1\\
	0,&\!\!\!\tr\rho_{A,B}^2=1,
\end{cases}\\
\check{\tau}_{G}^{AB}(|\psi\ra^{ABC})&=&{[h_\tau(\rho^A)h_\tau(\rho^B)]}^{\frac12}.
\eeax
If $|\psi\ra^{ABC}$ is pairwise entangled with respect to $AB$, then these quantities are not vanished, i.e., they are faithful. But for $|\psi_1\ra^{ABC}=|\psi\ra^{AB}|\psi\ra^C$, and $|\psi_2\ra^{ABC}$ is genuinely entangled with $\rho_1^A=\rho_2^A$, $\rho_1^B=\rho_2^B$, one has $\check{E}^{AB}(|\psi_1\ra^{ABC})=\check{E}^{AB}(|\psi_2\ra^{ABC})$ for any $\check{E}^{AB}\in\{\check{E}_{\min}^{AB},\check{\bar{E}}^{AB},\check{E}_{G}^{AB}\}$, where $\rho_i^{A,B}$ are the reduced state of $|\psi_i\ra^{ABC}$, $i=1$, $2$. That is, such measures lose the entanglement shared with the global system, but they are really well-defined PEMos.

\begin{figure}[ht]
	\centering
	\includegraphics[width=65mm]{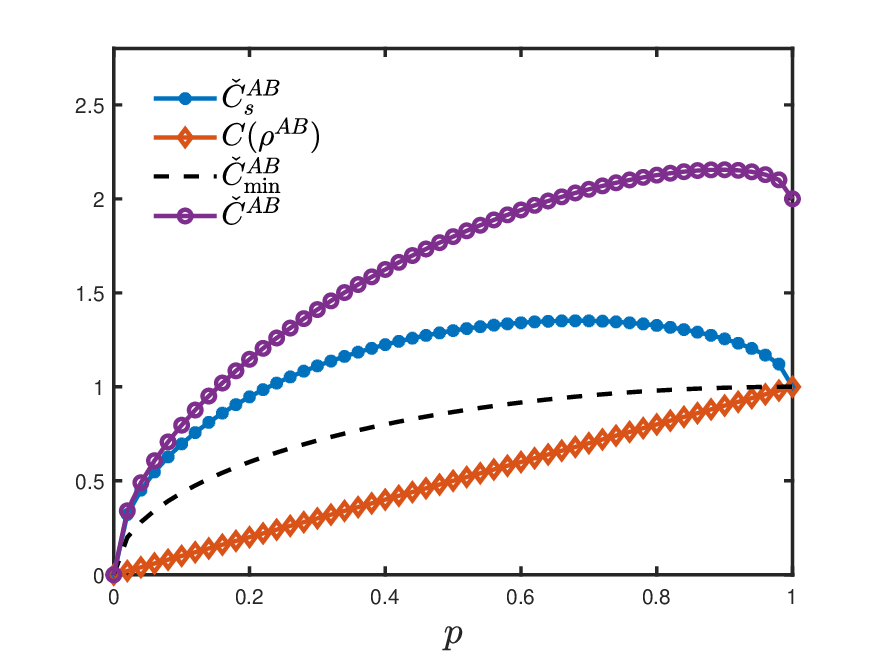}\\
	(a)
	
	\includegraphics[width=65mm]{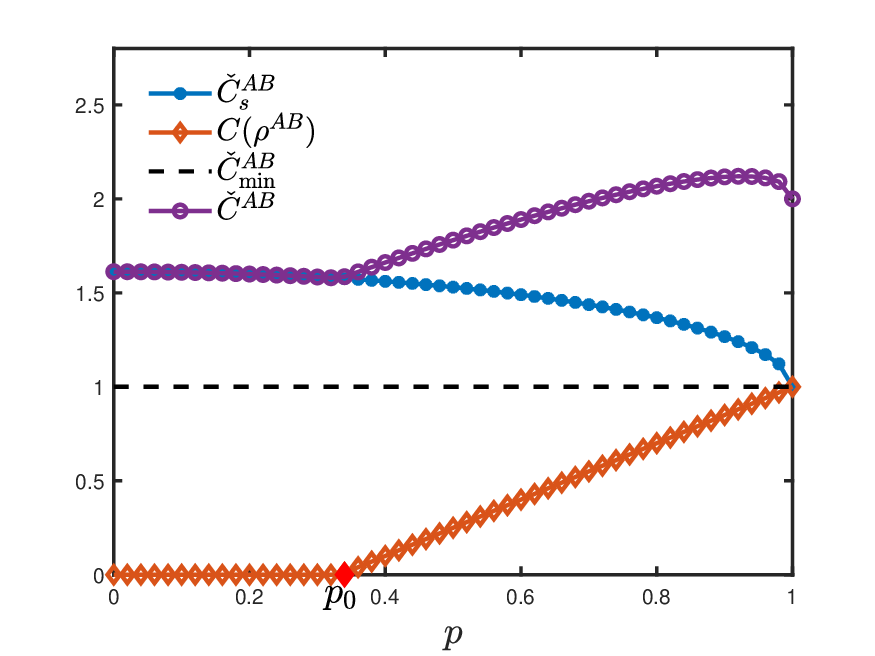}\\
	(b)
	\caption{\label{fig1} The PE of (a) $|\Psi_1\ra^{ABC}$, and
		(b) $|\Psi_2\ra^{ABC}$.
		We take $h_C(\rho)=\sqrt{2(1-\tr\rho^2)}$, and $C_g^{(3)}(|\psi\ra^{ABC})=\frac12\delta(|\psi\ra^{ABC})[h_C(\rho^A)+h_C(\rho^B)+h_C(\rho^C)]$.
		From the plot, we see that $\check{C}^{AB}$, $\check{C}_s^{AB}$, and $\check{C}_{\min}^{AB}$ are inequivalent to each other, where $\check{C}^{AB}(|\psi\ra^{ABC})=C(\rho^{AB})+C_g^{(3)}(|\psi\ra^{ABC})$, $\check{C}_s^{AB}(\rho^{AB})=C_g^{(3)}(|\Phi\ra^{ABC})$, and $\check{C}_{\min}^{AB}(|\psi\ra^{ABC})=\min\{C(A|BC), C(B|AC)\}$. Case (b) also reveals that the PE entanglement can decrease while the entanglement increases.}
\end{figure}

We take
\ba
&\check{\tau}_{ m}^{AB}(|\psi\ra^{ABC})=\tau(\rho^{AB})+\min\{h_\tau(A), h_\tau(B), h_\tau(C)\},\\
&\check{\tau}_s^{AB}(|\psi\ra^{ABC})=\min\{h_\tau(A), h_\tau(B), h_\tau(C)\},\nonumber
\ea
then $\check{\tau}_s^{AB}$ is a GEMo~\cite{Guo2024rp}. It turns out that
\beax
\mC_{A'B'}(|W\ra)=2/3<\mC_{A'B'}(|\text{\text{GHZ}}\ra)=1,
\eeax
\beax
\check{\tau}_{\rm m}^{AB}(|W\ra)=14/9>\check{\tau}_{\rm m}^{AB}(|\text{\text{GHZ}}\ra)=1,
\eeax
\beax
\check{\tau}_{\min}^{AB}(|W\ra)=8/9<\check{\tau}_{\min}^{AB}(|\text{\text{GHZ}}\ra)=1,
\eeax
\beax
\tau(\rho_{W}^{AB})=2/3>\tau(\rho_{\text{\text{GHZ}}}^{AB})=0,
\eeax
\beax
\check{\tau}_{s}^{AB}(|W\ra)=8/9<\check{\tau}_{s}^{AB}(|\text{GHZ}\ra)=1,
\eeax
where $\rho_{W}^{AB}=\tr_C|W\ra\la W|$, $\rho_{\text{GHZ}}^{AB}=\tr_C|\text{GHZ}\ra\la \text{GHZ}|$.
This also indicates that $\mC_{A'B'}$, $\check{\tau}_{\min}^{AB}$ and $\check{\tau}_{s}^{AB}$ are  ``proper'' PEMos but $\check{\tau}^{AB}_{m}$ is not proper, according to Ref.~\cite{Xie2021prl}.

Let
\bea
|\Psi_1\ra^{ABC}&=&\sqrt{p}|\phi^+\ra|0\ra^C+\sqrt{1-p}|\varphi\ra^A|\varphi\ra^B|1\ra^C, \label{Psi1}\\
|\Psi_2\ra^{ABC}&=&\frac{1}{2}(\sqrt{3p+1}|\phi^+\ra|0\ra^C+\sqrt{1-p}|\phi^-\ra|1\ra^C\nonumber\\
&&+\sqrt{1-p}|01\ra|2\ra^C
+\sqrt{1-p}|10\ra|3\ra^C),\label{Psi2}
\eea
where $|\phi^{\pm}\ra=\frac{1}{\sqrt2}(|00\ra\pm|11\ra)$.
We illustrate the relation between $\check{C}^{AB}$, $\check{C}_{\min}^{AB}$, $\check{C}_s^{AB}$, and $C(AB)$ in Fig.~\ref{fig1}.

This approach can be generalized to the $k$-PWEMo of $n$-party system case straightforwardly, $n>k\geqslant2$.
In general, we let
\bea\label{min-h-kpwem}
\check{E}_{\min}^{A_1A_2\cdots A_k}(|\psi\ra)=\min_{1\leqslant i\leqslant k}\min_{Z_i} h(A_iZ_i),
\eea
where $Z_i$ is contained in $\overline{A_i}$ but $Z_i\neq\overline{A_i}$ and $Z_i$ exclude at least one $A_j$, $j\neq i$, and the second minimum is taken over all possible $Z_i$ ($Z_i$ can be the empty set as well). We can also define
\begin{widetext}
\bea
\check{\bar{E}}^{A_1A_2\cdots A_k}(|\psi\ra)=\begin{cases}
	\frac12\sum\limits_{i=1}^kh(A_i), &\min\limits_{1\leqslant i\leqslant k}\min\limits_{Z_i} h(A_iZ_i)>0\\
	0,&\min\limits_{1\leqslant i\leqslant k} \min\limits_{Z_i}h(A_iZ_i)=0,
\end{cases}
\eea
 \bea
\check{E}_{G}^{A_1A_2\cdots A_k}(|\psi\ra)=\begin{cases}
\left[ \prod_{i=1}^kh(A_i)\right] ^{\frac1k}, &\min\limits_{1\leqslant i\leqslant k}\min\limits_{Z_i} h(A_iZ_i)>0\\
	0,&\min\limits_{1\leqslant i\leqslant k} \min\limits_{Z_i}h(A_iZ_i)=0,
\end{cases}
\eea
\end{widetext}
where $Z_i$ is taken as in Eq.~\eqref{min-h-kpwem}.

For any $\check{E}^{A_1A_2\cdots A_k}\in\{\check{E}_{\min}^{A_1A_2\cdots A_k}, \check{\bar{E}}^{A_1A_2\cdots A_k}, \check{E}_{G}^{A_1A_2\cdots A_k}\}$, $\check{E}^{A_1A_2\cdots A_k}(|\psi\ra)>0$ iff $|\psi\ra$ is S$k$-PWE since $\min\limits_{1\leqslant i\leqslant k}\min\limits_{Z_i} h(A_iZ_i)>0$ implies $|\psi\ra$ is S$k$-PWE.
Clearly, $\check{E}_{\min}^{A_1A_2\cdots A_k}$ and $\check{\bar{E}}^{A_1A_2\cdots A_k}$ are nonincreasing on average under LOCC. $\check{E}_{G}^{A_1A_2\cdots A_k}$ is also nonincreasing on average under LOCC since the geometric mean function $f =(\prod_i^nx_i)^{1/n}$ is a concave function~\cite{Boyd2004book} which guarantees that $\check{E}_{G}^{A_1A_2\cdots A_k}$ is a $k$-PWEMo according to Theorem 2 in Ref.~\cite{Vidal2000}. We thus conclude the following.

\begin{pro}
$\check{E}_{\min}^{A_1A_2\cdots A_k}$, $\check{\bar{E}}^{A_1A_2\cdots A_k}$, and $\check{E}_{G}^{A_1A_2\cdots A_k}$ are  faithful  S$k$-PWEMos, $k\geqslant3$.
\end{pro}

We would like to remark that, although $\check{E}_{\min}^{A_1A_2\cdots A_k}(\rho)>0$, $\check{\bar{E}}^{A_1A_2\cdots A_k}(\rho)>0$, and $\check{E}_{G}^{A_1A_2\cdots A_k}(\rho)>0$ if $\rho$ is G$k$-PWE, they are not G$k$-PWEMos since there exist non-G$k$-PWE state $\sigma$ such that
 $\check{E}_{\min}^{A_1A_2\cdots A_k}(\sigma)>0$, $\check{\bar{E}}^{A_1A_2\cdots A_k}(\sigma)>0$, and $\check{E}_{G}^{A_1A_2\cdots A_k}(\sigma)>0$. For example, we take $|\psi\ra^{ABCDEF}=|\text{GHZ}_4\ra|\phi^+\ra$, then $\check{E}_{\min}^{ABC}(|\psi\ra^{ABCDEF})>0$, $\check{\bar{E}}^{ABC}(|\psi\ra^{ABCDEF})>0$, and $\check{E}_{G}^{ABC}(|\psi\ra^{ABCDEF})>0$, but $|\psi\ra^{ABCDEF}$ is not a G$k$-PWE state.

\subsection{Distance based $k$-PWEMs}

The minimal distance from a given state to the set of some kind of separable states (biseparable states, fully separable states, $k$-separable states, etc.) is a popular method applied in quantifying various entanglement, such as the relative entropy of entanglement~\cite{Vedral1998pra,Piani2009prl}, the geometric measure of entanglement~\cite{Shimony95,Cao2007jpa,Streltsov2010njp}, and so on. We show below that this approach is also valid for the $k$-partitewise entanglement.

We denote the set of all $k$-partitewise separable states in $\mS^{A_1A_2\cdots A_n}$ by $\mS_{pws}^{A_1A_2\cdots A_k}$, and the set of $k$-partitewise separable pure states in $\mH^{A_1A_2\cdots A_n}$ by $\mP_{pws}^{A_1A_2\cdots A_k}$.
We let
\bea
\check{S}_r^{A_1A_2\cdots A_k}(\rho)&=&\min\limits_{\sigma\in\mS_{pws}^{A_1A_2\cdots A_k}}S(\rho\|\sigma),\\
\check{E}_{\rm G}^{A_1A_2\cdots A_k}(|\psi\ra)&=&1-\max\limits_{|\phi\ra\in\mP_{pws}^{A_1A_2\cdots A_k}}|\la\psi|\phi\ra|^2,
\eea
where $S(\rho\|\sigma)=\tr[\rho(\log_2\rho-\log_2\sigma)]$ is the relative entropy.
Since $\mS_{pws}^{A_1A_2\cdots A_k}$ is a convex and compact set, $\check{S}_r^{A_1A_2\cdots A_k}$ is a $k$-PWEMo according to the argument in~\cite{Piani2009prl}. $\check{E}_{\rm G}^{A_1A_2\cdots A_k}$ is a $k$-PWEM since the geometric measure of entanglement was shown to be non-increasing under LOCC~\cite{Cao2007jpa,Streltsov2010njp}.


\section{Partitewise entanglement extension}


The $k$-partitewise entanglement discuss the shared entanglement of fixed subsystem in a given global system. In this section, we consider another issue: For a given state such that we do not know the global system it lived in, whether there exists a larger global state such that it is $k$-PWE with respect to the given state, and moreover, how we can quantify such a capability and what the relation is between this capability and the $k$-PWE of the global state.

We start with the pairwise entanglement case. Let $\rho^{AB}$ be a mixed state. We assume for a moment that $\rho^{AB}$ is separable. If $\rho^A$ or $\rho^B$ is a pure state, then any extension of $\rho^{AB}$ must be pairwise separable with respect to $AB$. If $\min\{r(\rho^A), r(\rho^B)\}\geqslant2$, we let $\rho^{AB}=\sum_ip_i|\psi_i\ra\la\psi_i|^A\ot|\psi_i\ra\la\psi_i|^B$. Taking
\bea \label{purification1}
|\Psi\ra^{ABC}=\sum_i\sqrt{p_i}|\psi_i\ra^A|\psi_i\ra^B|i\ra^C
\eea
with $\{|i\ra^C\}$ an orthogonal set of $\mH^C$,
it follows that $|\Psi\ra^{ABC}$ is genuinely entangled, which leads to $\rho^{AB}$ being pairwise entangled.
If $\rho^{AB}$ is entangled, it is of course pairwise entangled. For any mixed entangled state $\rho^{AB}$, let $\rho^{AB}=\sum_jq_j|\psi_j\ra\la\psi_j|^{AB}$ be the spectral decomposition,
then
\bea\label{purification2}
|\Phi\ra^{ABC}=\sum_j\sqrt{q_j}|\psi_j\ra^{AB}|j\ra^C
\eea
is genuinely entangled. That is, any bipartite mixed state that contains no pure reduced state can be extended into a genuine entangled state in a larger tripartite system. Here, mixed state $\rho^{AB}$ contains no pure reduced state refers to both $\rho^A$ and $\rho^B$ are mixed, $\rho^{A,B}=\tr_{B,A}\rho^{AB}$. This motivates us to give the following concepts.

\begin{definition}
A state $\rho\in\mS^{A_1A_2\cdots A_k}$ is {\it $k$-partitewise entanglement extendable ($k$-PWEE)} if there exists a $k$-PWE state $\rho^{A_1A_2\cdots A_l}$ with respect to $A_1A_2\cdots A_k$ such that $\rho=\tr_{A_{k+1}\cdots A_l}\rho^{A_1A_2\cdots A_l}$ and $\rho^{A_1A_2\cdots A_l}\neq \rho^{A_1A_2\cdots A_k}\ot\rho^{A_{k+1}\cdots A_l}$, $k<l$. In such case, $\rho^{A_1A_2\cdots A_l}$ is called a $k$-partitewise entanglement extension of $\rho$. In particular, $\rho$ is {\it strongly $k$-partitewise entanglement extendable (S$k$-PWEE)} if there exists a genuinely entangled state $\rho^{A_1A_2\cdots A_l}$ such that $\rho=\tr_{A_{k+1}\cdots A_l}\rho^{A_1A_2\cdots A_l}$, $k<l$, and $\rho^{A_1A_2\cdots A_l}$ is called a strong $k$-partitewise entanglement extension of $\rho$.
\end{definition}

 Clearly, any pure state $|\psi\ra^{A_1A_2\cdots A_k}$ is not $k$-partitewise entanglement extendable. The methods in Eqs.~\eqref{purification1} and \eqref{purification2} are also valid for the multipartite case (notice that, up to local unitary operation, $|\Psi\ra^{ABC}$ in Eq.~\eqref{purification1} coincides with $|\Phi\ra^{ABC}$ in Eq.~\eqref{purification2} indeed). This leads to the following theorem.

\begin{theorem}
$\rho\in\mS^{A_1A_2\cdots A_k}$ is $k$-PWEE iff $\rho$ has at least one bipartite mixed reduced state that contains two mixed reduced states (if $k=2$, $\rho$ is a mixed state that contains two mixed reduced states), and it is S$k$-PWEE iff it is a mixed state that contains no pure reduced state.
\end{theorem}

For example, $\rho^{ABC}=\rho^{AB}\ot\rho^{C}$ with $r(\rho^{AB})\geqslant 2$ and $r(\rho^{A,B})\geqslant 2$ is three-partitewise entanglement extendable; mixed state $\varrho^{ABC}$ with $r(\rho^{AB,AC,BC})\geqslant 2$ and $r(\varrho^{A,B,C})\geqslant 2$ is not only $3$-PWEE and but also S$3$-PWEE. In fact, if $\varrho^{ABC}$ is mixed with $r(\rho^{AB,AC,BC})\geqslant 2$ and $r(\varrho^{A,B,C})\geqslant 2$, then any purification of $\varrho^{ABC}$, $|\Psi\ra^{ABCD}$, has no pure reduced state, so $|\Psi\ra^{ABCD}$ is genuinely entangled.

\subsection{Measure of partitewise entanglement extensibility}

The argument above indicates that, not only the (fully) separable state is partitewise entanglement extendable but also the entangled state. However, the
partitewise entanglement extension is limited. For example, if the pairwise entanglement extension of $\rho^{AB}$ is the generalized \text{GHZ} state, then $\rho^{AB}$ cannot be entangled. This motivates us to provide a measure of the pairwise entanglement {\it extensibility}
\bea \label{extensiblity}
E_{\text{ext}}(\rho^{AB})=
\begin{cases}
E_g^{(3)}(|\Phi\ra^{ABC}),& \!\!\!\text{$\rho$ is PE extendable}~~~\\
0, &\!\!\!\text{otherwise},
\end{cases}
\eea
where $|\Phi\ra^{ABC}$ is defined as in Eq.~\eqref{purification2}, $E_g^{(3)}$ is some genuine entanglement measure. Intuitively, $E_{\text{ext}}$ depends on not only eigenvalues of $\rho^{AB}$, but also that of $\rho^{A,B}$, in general.

If the partitewise entanglement extension is an absolutely maximally entangled state, then, generally, it has maximal entanglement extensibility. For example,
the entanglement extensibility of $\rho^{AB}=\frac12(|00\ra\la00|+|11\ra\la11|)$ is $E_g^{(3)}(|\text{GHZ}\ra)$, where $|\text{GHZ}\ra$ is an absolutely maximally entangled state, and that of $\varrho^{AB}=\frac14I_4$ equals to $E_g^{(3)}(|\Phi_5\ra)$, where $|\Phi_5\ra=\frac{1}{\sqrt8}(-|00000\ra+|01111\ra-|10011\ra+|11100\ra+|00110\ra+|01001\ra+|10101\ra+|11010\ra)$ is an absolutely maximally entangled state in a five-qubit system~\cite{Goyeneche2014pra} and $I_4$ denotes the identity operator on the two-qubit space.
We depict the relation between $E_{\text{ext}}(AB)$, the linear entropy of $\rho^{AB,A,B}$, and the entanglement $E(AB)$ in Figs.~\ref{fig2} and~\ref{fig3}.

\begin{figure}[htp]
	\centering
	\includegraphics[width=56mm]{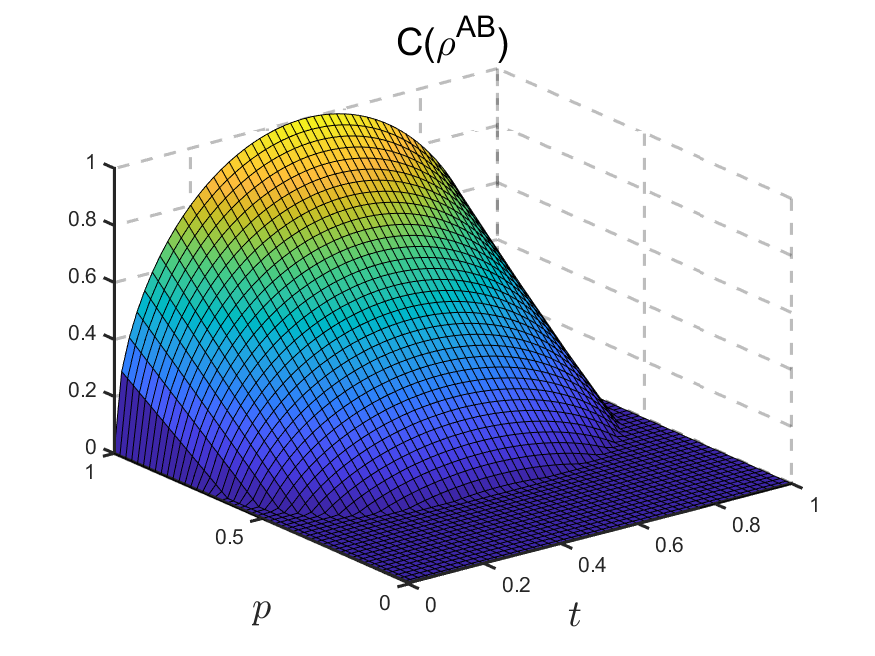}\\
	\includegraphics[width=56mm]{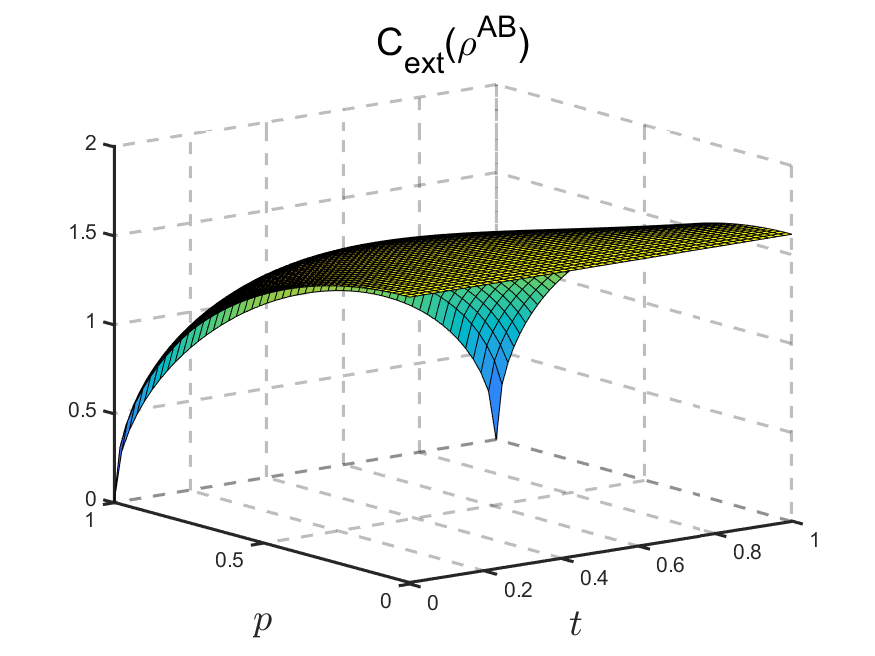}\\
	\includegraphics[width=56mm]{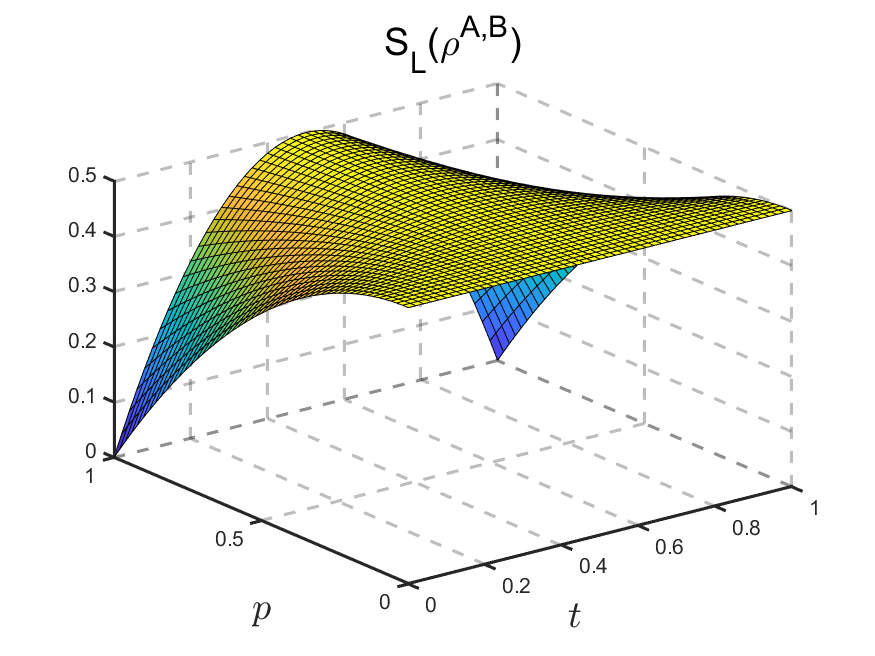}\\
	\includegraphics[width=56mm]{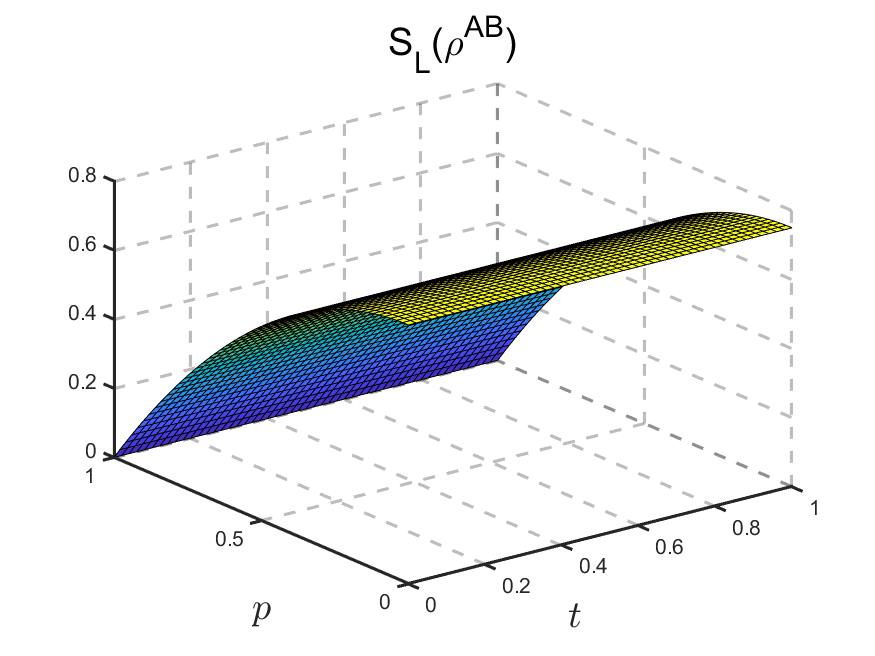}
	\vspace{-2mm}
	\caption{\label{fig2} The PE extensibility of $\rho^{AB}=p|\phi\ra\la\phi|+(1-p)\frac{I_4}{4}$ with $|\phi\ra=\sqrt{t}|00\ra+\sqrt{1-t}|11\ra$, $0\leqslant p, t\leqslant 1$. Here, $h=h_C$, $C_{\text{ext}}(\rho^{AB})=C_g^{(3)}(|\Phi\ra^{ABC})$. For any fixed $p<0.5$, $E(AB)\nearrow$ when $E_{\text{ext}}\nearrow$, but for any fixed $0<t<1$, $E(AB)\searrow$ when $E_{\text{ext}}\nearrow$. In addition, $E_{\text{ext}}\nearrow$ whenever  $S_L^{AB,A,B}\nearrow$,  where $S_L(\rho)=1-\tr\rho^2$ denotes the purity of $\rho$.
	}
\end{figure}

\begin{figure}[htp]
	\centering
	\includegraphics[width=65mm]{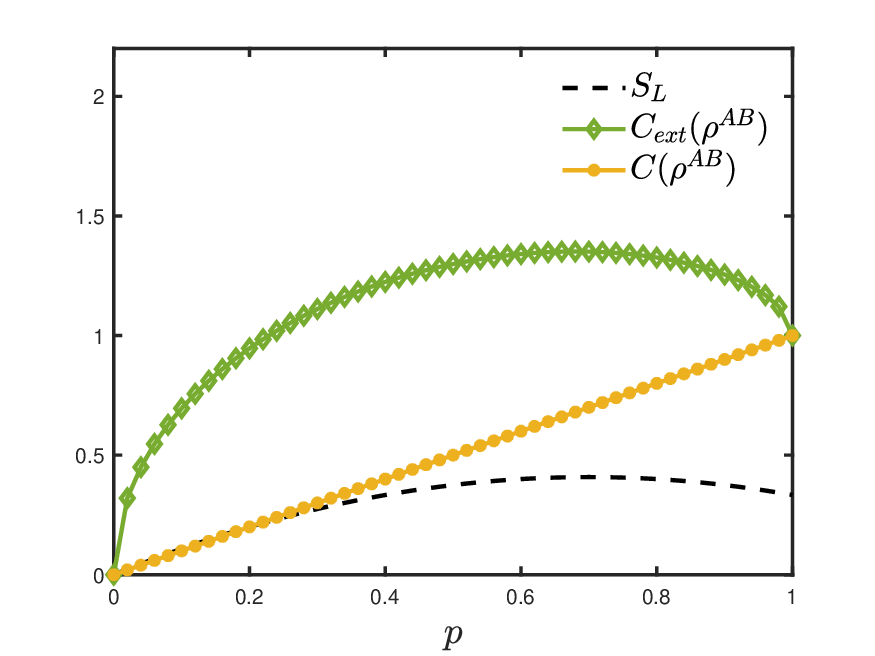}
	
	(a)
	
	\includegraphics[width=65mm]{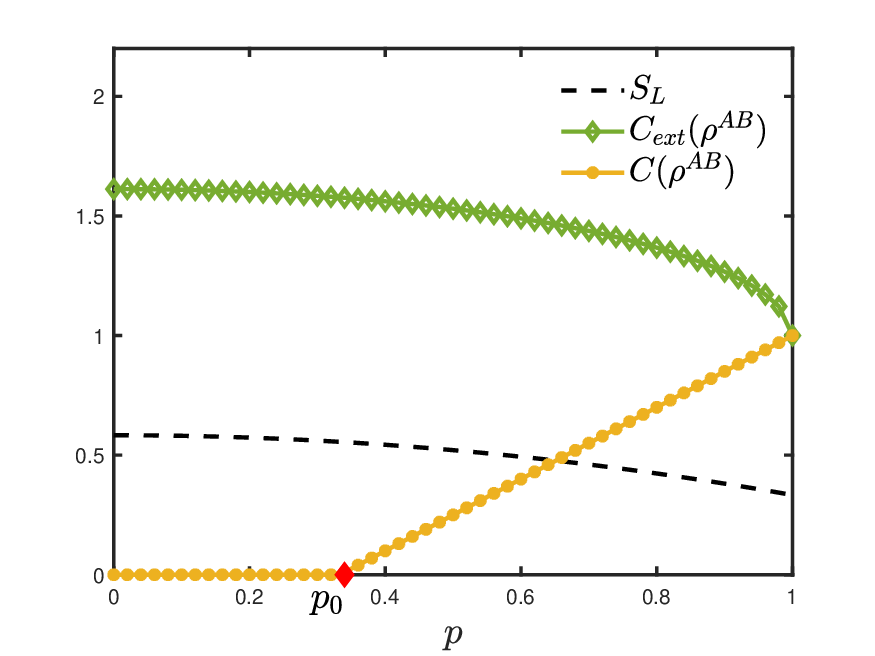}
	
	(b)
	\caption{\label{fig3} The PE extensibility of (a) $\rho^{AB}=p|\Phi^+\ra\la\Phi^+|+(1-p)\sigma^{AB}$, $\sigma^{AB}=|\varphi\ra\la\varphi|^{A}\ot|\varphi\ra\la\varphi|^{B}$, $|\varphi\ra^{A}=\frac{1}{\sqrt{2}}(|0\ra+|1\ra)$, $|\varphi\ra^{B}=|0\ra$, and
		(b) $\rho^{AB}=p|\Phi^+\ra\la\Phi^+|+(1-p)\frac{I_4}{4}$. Here, $h=h_C$, $C_{\text{ext}}(\rho^{AB})=C_g^{(3)}(|\Phi\ra^{ABC})$, $|\Phi^+\ra=\frac{1}{\sqrt2}(|00\ra+|11\ra)$, $p_0=\frac13$. In case (a), $E_{\text{ext}}$ is monotonically decreasing convergent to 1 whenever $E(AB)$ is monotonically increasing convergent to 1 for $p>0.681$. But for $p<0.681$, both of them are monotonically increasing. At the same time, $S_L:=\frac13(S_L(AB)+S_L(A)+S_L(B))$ is increasing whenever $p<0.7$ and decreasing whenever $p>0.7$. For case (b), the relation is more clear: $\rho^{AB}$ is entangled when $p>\frac13$ and $E(AB)$ is monotonically increasing for $p>\frac13$, while $S_L$ is monotonically decreasing. In both cases, the shape of line for $E_{\text{ext}}$ and that for $S_L$ keep almost the same variation tendency.}
\end{figure}

For any given state $\rho^{A_1A_2\cdots A_k}$, we denote by $\mS_{\text{ext}}(\rho^{A_1A_2\cdots A_k})$ the set of all the $k$-paritewise entanglement extensions of $\rho^{A_1A_2\cdots A_k}$ and by $\mS_{\text{sext}}(\rho^{A_1A_2\cdots A_k})$ the set of all the strong $k$-paritewise entanglement extensions of $\rho^{A_1A_2\cdots A_k}$. It is easy to see that
\beax
\mS_{\text{sext}}(\rho^{AB})=\mS_{\text{ext}}(\rho^{AB}),
\eeax
i.e., $\rho^{AB}$ is pairwise entanglement extendable iff it is strongly pairwise entanglement extendable. But
\beax
\mS_{\text{sgext}}(\rho^{A_1A_2\cdots A_k})\subsetneq\mS_{\text{ext}}(\rho^{A_1A_2\cdots A_k}),
\eeax
whenever $k\geqslant3$. It turns out that both $\mS_{\text{ext}}(\rho^{A_1A_2\cdots A_k})$ and $\mS_{\text{sext}}(\rho^{A_1A_2\cdots A_k})$ are convex sets with all the purification as in Eqs.~\eqref{purification1} and~\eqref{purification2} are the extreme points. But $\mS_{\text{ext}}(\rho^{A_1A_2\cdots A_k})$ is not the convex hull of these points. For example, $\rho_p=p\rho^{AB}\ot\rho^C+(1-p)|\Phi\ra\la\Phi|\in \mS_{\text{ext}}(\rho^{AB})$, $0<p<1$, but $\rho_p$ is not a convex combination of these extreme points. In what follows, we show that $E_{\text{ext}}$ of $\rho^{AB}$ achieves the maximal value over all its strong pairwise entanglement extensions.

\begin{theorem}\label{th2}
Let $\rho^{AB}$ be a mixed state in $\mS^{AB}$, and $r(\rho^{A,B})>1$. Then
\bea
E_{\text{ext}}(\rho^{AB})=\max\limits_{\rho\in\mS_{\text{sext}}(\rho^{AB})} E_g^{(3)}(\rho).
\eea
\end{theorem}

\begin{proof}
We let $\rho^{AB}=\sum_j^mq_j|\psi_j\ra\la\psi_j|^{AB}$ be its spectral decomposition, and let $\alpha$ be a proper subset of $\{1, 2, \dots, m\}$; $\alpha$ includes at least two elements,
\beax
p_\alpha\rho_1^{AB}=\sum_{j\in\alpha}q_j|\psi_j\ra\la\psi_j|^{AB},
\eeax
\beax
(1-p_\alpha)\rho_2^{AB}=\sum_{j\in\overline{\alpha}}q_j|\psi_j\ra\la\psi_j|^{AB}
\eeax
where $\overline{\alpha}$ denotes the complementary set of $\alpha$ in $\{1, 2, \dots, m\}$, $p_\alpha=\sum_{j\in\alpha}q_j$, $q_j>0$. Let $|\Phi_\alpha\ra\in\mH^{ABC}$ be the purification of $\rho_1^{AB}$; then
\beax
\rho_\alpha=(1-p_\alpha)\rho_2^{AB}\ot\rho^C+p_\alpha|\Phi_\alpha\ra\la\Phi_\alpha|\in\mS_{\text{sext}}(\rho^{AB})
\eeax
for any $\rho^C\in\mS^C$. If $\rho\in\mS_{\text{sext}}(\rho^{AB})$, there are three cases: (i) $\rho=|\Phi\ra\la\Phi|^{ABC}$, where $|\Phi\ra^{ABC}$ is defined as in Eq.~\eqref{purification2}; (ii) $\rho=\rho_p=p\rho^{AB}\ot\rho^C+(1-p)|\Phi\ra\la\Phi|^{ABC}$; and (iii) $\rho=\rho_\alpha$. It is clear that case (i) reaches the maximal value of $E_g^{(3)}(\rho)$, which completes the proof  (for the special case $m=2$, $\rho_\alpha=|\Phi_\alpha\ra\la\Phi_\alpha|=|\Phi\ra\la\Phi|^{ABC}$ and there are only two cases for $\rho$).
\end{proof}

That is, $E_{\text{ext}}(\rho^{AB})$ is uniquely determined by $\rho^{AB}$ itself since $E_g^{(3)}$ depends only on eigenvalues of $\rho^{AB}$ and that of $\rho^{A,B}$ generally. Namely, although the global system is unknown, up to local unitary operation, the maximal pairwise entanglement extension is exactly decided by the given state. This also manifests that the state supposition is the nature of the entanglement.

We now consider the extensibility of $\rho^{ABC}$. If $\rho^{ABC}=\rho^{AB}\ot|\psi\ra\la\psi|^C$ with $r(\rho^{AB,A,B})\geqslant2$, it is extendable. Namely, there exits genuinely entangled $|\Phi\ra^{ABD}$ such that $\tr_D|\Phi\ra\la\Phi|=\rho^{ABC}$.
Then $|\Phi\ra^{ABD}|\psi\ra^C$ is the three-partitewise entanglement extension of $\rho^{ABC}$. If $\rho^{AB}=\rho^A\ot\rho^B$ additionally,
then there exist purification of $\rho^A$ and $\rho^B$, $|\phi\ra^{AD}$ and $|\phi\ra^{BE}$, respectively. But
$|\phi\ra^{AD}|\phi\ra^{BE}|\psi\ra^C$ is not a three-partitewise entanglement extension of $\rho^{ABC}$ since $|\phi\ra^{AD}|\phi\ra^{BE}|\psi\ra^C$ is not a three-partitewise entangled state with respect to $ABC$.
If $r(\rho^{ABC,AB,AC,BC,A,B,C})\geqslant2$, there are three cases: (a) $\rho^{ABC}=\rho^A\ot\rho^B\ot\rho^C$; (b)  $\rho^{ABC}=\rho^A\ot\rho^{BC}$ (or $\rho^{ABC}=\rho^{AB}\ot\rho^{C}$, or $\rho^{ABC}=\rho^B\ot\rho^{AC}$); and (c) $\rho^{ABC}$ is not a product state. It results in different three-partitewise entanglement extensions. Also note that they always have a common three-partitewise entanglement extension $|\Phi\ra^{ABCD}$ that is a genuinely entangled state. Thus, when $k\geq3$, the entanglement extension is not unique (up to local unitary operation). So, with the notations as in Eq.~\eqref{kpwem}, we define
\bea \label{k-ent-ext0}
E_{\text{ext}}^{\psi}(\rho^{A_1A_2\cdots A_k})=
	\sum\limits_{X_s\in\varUpsilon^{X_1X_2\cdots X_l}}E_g^{(t_s)}(|\psi\ra^{X_s}),
\eea
if there exits a $k$-partitewise entanglement extension $|\psi\ra=|\psi\ra^{X_1}|\psi\ra^{X_2}\cdots|\psi\ra^{X_l}$, $l<k$.
Namely,
\begin{widetext}
\bea \label{k-ent-ext}
E_{\text{ext}}(\rho^{A_1A_2\cdots A_k})=
\begin{cases}
E_{\text{ext}}^{\psi}(\rho^{A_1A_2\cdots A_k}), &\text{$\rho^{A_1A_2\cdots A_k}$ is $k$-PWEE}\\
0, & \text{otherwise},
\end{cases}
\eea
\end{widetext}
with some abuse of notation.
 And in particular, we define
\bea\label{m-k-ent}
E_{\text{ext}}^{\Phi}(\rho^{A_1A_2\cdots A_k})=E_g^{(k+1)}(|\Phi\ra^{A_1A_2\cdots A_kA_{k+1}}),
\eea
if $|\Phi\ra^{A_1A_2\cdots A_kA_{k+1}}$ is a strong $k$-partitewise entanglement extension of $\rho^{A_1A_2\cdots A_k}$. Analogous to Theorem~\ref{th2}, one can derive
\bea E_{\text{ext}}^{\Phi}(\rho^{A_1A_2\cdots A_k})=\max\limits_{\rho\in\mS_{\text{sext}}(\rho^{A_1A_2\cdots A_k})}E_g^{(k+1)}(\rho).
\eea

\subsection{PWE and PWE extensibility}

For any S$k$-PWE state $|\psi\ra$, with the notations as in Eqs.~\eqref{kpwem} and~\eqref{skpwem}, it is clear that
\beax
&&\check{E}^{A_1A_2\cdots A_k}(|\psi\ra)\\
&=&E^{(k)}(A_1A_2\cdots A_k)+E_g^{(t_X)}(|\psi\ra^{X})\\
&=&\!E^{(k)}(A_1A_2\cdots A_k)+E_{\text{ext}}(A_1A_2\cdots A_k).
\eeax
In particular, for any tripartite pure state $|\psi\ra^{ABC}$,
\beax
\check{E}^{AB}(|\psi\ra^{ABC})=E(AB)+E_{\text{ext}}(AB),
\eeax since $\check{E}_s^{AB}(|\psi\ra^{ABC})=E_{\text{ext}}(\rho^{AB})$ whenever one choose the same $E_g^{(3)}$ in $\check{E}_s^{AB}$ and $E_{\text{ext}}$. Together with Eqs.~\eqref{pem}, \eqref{extensiblity}, \eqref{k-ent-ext0}, \eqref{k-ent-ext} and \eqref{m-k-ent}, we have
\bea
\check{E}^{A_1A_2\cdots A_k}(|\psi\ra)\!=\!E^{(k)}(A_1A_2\cdots A_k)+E_{\text{ext}}^{\psi}(A_1A_2\cdots A_k),\nonumber
\\
\eea
with some abuse of notation (here $E_{\text{ext}}^{\psi}:=0$ if $\rho^{A_1A_2\cdots A_k}$ is not $k$-partitewise entanglement extendable). In other words, $E_{\text{ext}}$ is just the ``global entanglement'' shared by $A_1A_2\cdots A_k$ in its extension $|\psi\ra$ under the measure $\check{E}^{A_1A_2\cdots A_k}$.

We now discuss the relation between $E^{(k)}(A_1A_2\cdots A_k)$ and $E_{\text{ext}}^{\psi}(A_1A_2\cdots A_k)$ again. We begin with the case of $k=2$ for clarity. If $\rho^{AB}$ is maximally entangled, then it is pure~\cite{Guo2020pra}, this results in $E_{\text{ext}}(AB)=0$. Conversely, if a three-qubit state $|\psi\ra^{ABC}$ is absolutely maximally entangled, i.e., $|\psi\ra^{ABC}$ is the \text{GHZ} state, then $E(AB)=0$. In addition, the \text{GHZ} state is regarded more entangled than the $W$ state~\cite{Xie2021prl}. So this seemingly indicates that, $E(AB)$ and $E_{\text{ext}}$ are complementary to each other. But from Figs.~\ref{fig2} and \ref{fig3}, there exist states such that both $E(AB)$ and $E_{\text{ext}}$ can increase simultaneously, and states such that $E(AB)$ creases with $E_{\text{ext}}$ decreases. Figure~\ref{fig4} implies that there also exist states such that $E(AB)$ decreases with $E_{\text{ext}}$ increases, and states such that $E(AB)$ remain invariant with $E_{\text{ext}}$ increases.

\begin{figure}[htp]
	\centering
	\includegraphics[width=65mm]{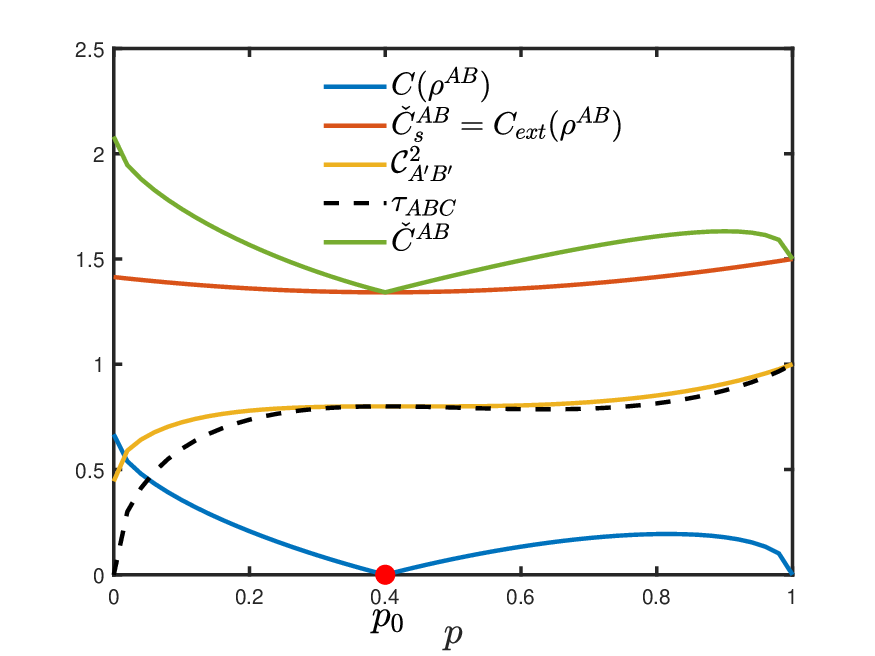}
	\caption{\label{fig4} The pairwise entanglement, entanglement, and the pairwise entanglement extensibility of $|\Psi_3\ra^{ABC}=\sqrt{p}|\text{GHZ}\ra+\sqrt{1-p}|W\ra$ under $\check{C}^{AB}$ and $\mC_{A'B'}^2$. $p_0=0.4$.}
\end{figure}

Another issue is: what is the supremum of $E(AB)+E_{\text{ext}}(AB)$? It seems a difficult task but
\bea
E(AB)+E_{\text{ext}}(AB)<2
\eea
for any bipartite state with any normalized measures $E(AB)$ and $E_{\text{ext}}(AB)$ since they cannot be maximally entangled at the same time. Here, an entanglement measure $E$ is called normalized if $|E|\leqslant1$ for any state and $E=1$ for the maximally entangled states. Besides Figs.~\ref{fig2}--\ref{fig4}, we list these quantities in Table~\ref{tab:table0} as a futher illustration of the relation.

\begin{table*}
\caption{\label{tab:table0} The relations between $E(AB)$, $E_{\text{ext}}$ and $\check{E}(AB)$. }	
\begin{ruledtabular}
\begin{tabular}{ccccc}
PEM	                &	state          & $E(AB)$ & $E_{\text{ext}}(AB)$ & $\check{E}^{AB}$\\ \colrule
$\mC^2_{A'B'}$		& $|W\ra$          &  $4/9$       &  0 &$4/9$\\			
$\mC^2_{A'B'}$		& $|\text{GHZ}\ra$        &   $0$     &  $1$ &$1$\\	
$\mC^2_{A'B'}$		& $|\Psi_3\ra$ in Fig.~\ref{fig4} & $p\in[0, 0.4]\searrow$, $[0.4, 0.816]\nearrow$
& $p\in[0, 0.4]\nearrow$, $[0.4, 0.636]\searrow$   &  $p\in[0, 0.4]\nearrow$, $[0.4, 0.469]\searrow$ \\
		&    & $[0.816, 1]\searrow$
&  $[0.636, 1]\nearrow$  & $[0.469, 1]\nearrow$ \\
&&&&  0.8 at $p=0.4$, 1 at $p=1$	\\			
$\check{\tau}^{AB}\footnotemark[1]$		&  $|W\ra$          &   $2/3$      & $4/3$  &$2$\\
$\check{\tau}^{AB}$		&  $|\text{GHZ}\ra$          & $0$        &$3/2$   &	$3/2$\\
$\check{C}^{AB}$		&  $|\Psi_1\ra$ in Eq.\!~\eqref{Psi1}        &  $p\in[0, 1] \nearrow $     &  $p\in[0, 0.681] \nearrow $, $[0.681, 1] \searrow $  & $p\in[0, 0.849] \nearrow $, $[0.849, 1] \searrow$\\
&&&&  2.0807 at $p=0.849$\\
$\check{C}^{AB}$		&  $|\Psi_2\ra$ in Eq.\!~\eqref{Psi2}      &  $p\in[0,\frac13] \equiv0, [\frac13, 1] \nearrow $      & $p\in[0, 1] \searrow $  & $p\in[0, 0.926]\nearrow $, $[0.926, 1] \searrow $\\
&   &&& 2.1201 at $p=0.926$\\
$\check{C}^{AB}$ 	& $|\Psi_3\ra$ in Fig.~\ref{fig4}    & $p\in[0, 0.4]\searrow$, $[0.4, 0.816]\nearrow$
& $p\in[0, 0.4]\searrow$    &$p\in[0, 0.4]\searrow$, $[0.4, 0.901]\nearrow$ \\
	&     & $[0.816, 1]\searrow$
&  $[0.4, 1]\nearrow$    &$[0.901, 1]\searrow$ \\
&&&&   1.6317 at $p=0.901$\\
&&&&  2.0809 at $p=0$
\end{tabular}
\end{ruledtabular}
	\footnotetext[1]{$\check{\tau}^{AB}(|\psi\ra^{ABC}):=\tau(AB)+\tau_g^{(3)}(ABC)$, where $\tau_g^{(3)}(|\psi\ra^{ABC})=\frac12\delta(|\psi\ra^{ABC})(h_\tau(A)+h_\tau(B)+h_\tau(C))$.}
\end{table*}

Let $\varepsilon^{AB}$ be a LOCC on $\mS^{AB}$. For any given $\rho\in\mS^{ABC}$, let $\varepsilon(\rho)=\varepsilon^{AB}\ot\mathbbm{1}^C(\rho)=\sigma\in\mS^{ABC}$. Then
\bea \label{ext-var}
E_{\text{ext}}(\sigma^{AB})-E_{\text{ext}}(\rho^{AB})\leqslant E(\rho^{AB})-E(\sigma^{AB}),
\eea
since $\check{E}^{AB}(\sigma)\leqslant\check{E}^{AB}(\rho)$. In other words, if $E_{\text{ext}}(\sigma^{AB})\geqslant E_{\text{ext}}(\rho^{AB})$, the increment of the entanglement extensibility of $AB$ is not exceeding the decrement of entanglement in $AB$ under LOCC. Note that it may happen that $E_{\text{ext}}(\sigma^{AB})<E_{\text{ext}}(\rho^{AB})$. For example, we take $\rho^{AB}$ as the reduced state of the $W$ state, i.e., $\rho^{AB}=\frac13(|10\ra\la10|+|10\ra\la01|+|01\ra\la10|+|01\ra\la01|+|00\ra\la00|)$, $\varepsilon^{AB}(\cdot)=\sum_{i=1}^4M_i(\cdot)M_i^\dag$ with $M_1=|00\ra\la00|$, $M_2=|00\ra\la01|$, $M_3=|00\ra\la10|$, and $M_4=|00\ra\la11|$. It turns out that $\sigma^{AB}=|00\ra\la00|$ is a pure state, which leads to
\bea E_{\text{ext}}(\sigma^{AB})=0<E_{\text{ext}}(\rho^{AB}).
\eea
If we take $M_1=|00\ra\la00|$, $M_2=|01\ra\la01|$, $M_3=|10\ra\la10|$, and $M_4=|11\ra\la11|$, then $\sigma^{AB}=\frac13(|00\ra\la00|+|01\ra\la01|+|10\ra\la10|)$. If we take $E_g^{(3)}=\tau_g^{(3)}$ as in Table~\ref{tab:table0}, then
\bea
E_{\text{ext}}(\sigma^{AB})=\frac{14}{9}>E_{\text{ext}}(\rho^{AB})=\frac43.
\eea


\section{Conclusion and discussion}


We have established an outline of the partitewise entanglement theory. It differs from both the $k$ entanglement and the $k$-partite entanglement. Indeed it is closely related to the structure of the multipartite entanglement (especially the genuine entanglement) which provides tools for the classification of genuine entanglement. We also proposed the concept, termed PWE extensibility, which reflects the shared capabilities of entanglement in a larger global system, but behaves contrary to that of the general entanglement measure in some sense. For any given state, the PWE extensibility together with the entanglement in it is just the $k$-partitewise entanglement in the extension state, and the two parts are limited by each other to some degree.

We have also proposed three types of $k$-PWEMs, in which the two former ones include many cases since we can choose any possible reduced function. Of course, we can quantify the amount of the partite-wise entanglement from other perspectives. For instance, observing that, if the reduced state $\rho^{AB}=\tr_C|\psi\ra\la\psi|^{ABC}$
is not a product state, i.e., $\rho^{AB}\neq\rho^A\ot\rho^B$, there exists pairwise entanglement in $AB$. On the other hand, $\rho^{AB}=\rho^A\ot\rho^B$ iff $I(A:B)=S(A)+S(B)-S(AB)=0$, where $I(A:B)$ is the mutual information of $\rho^{AB}$. Thus $I(A:B)$ measure the pairwise entanglement (the $n$-partite case follows as consequence via the multiparty mutual information~\cite{Watanabe1960,Kumar2017pra}), but it may increase under LOCC.

Furthermore the entanglement extensibility tells us that, even though a given state is separable, it may be entangled with another party in a larger system. So when Alice and Bob share a mixed state (separable or entangled) that contains no pure reduced state, they can cooperate with Charlie (as a environment) by sharing a three-party genuinely entangled state. The maximal sharability is limited by $E_{\text{ext}}$. What is more, the pure state seems more secure than the mixed state in the sense that Charlie may eavesdrop some information shared by Alice and Bob from the global state extended. So our contributions shed new light on the feature of the entanglement (superposition) as a resource, which need further in-depth research.


\begin{acknowledgements}
We are grateful to the anonymous referees for valuable comments and suggestions, which have contributed to clarifying and improving the paper. This work is supported by the National Natural Science Foundation of China under Grant No.~12471434 and No.~11971277, the Program for Young Talents of Science and Technology in Universities of Inner Mongolia Autonomous Region under Grant No. NJYT25010, and the High-Level Talent Research Start-up Fund of Inner Mongolia University under Grant No. 10000-2311210/049.
\end{acknowledgements}


	

\end{document}